\documentclass{article}
\usepackage{spconf,amsmath,epsfig,amsfonts,amssymb}
\usepackage{setspace}
\usepackage[misc]{ifsym}

\begin{document}\sloppy
\small
\begin{spacing}{0.88}
% Example definitions.
% --------------------
\def\x{{\mathbf x}}
\def\L{{\cal L}}

% Title.
% ------
\title{Herding Effect based Attention for Personalized Time-Sync Video Recommendation}
%
% Single address.
% ---------------
\name{Wenmian Yang$^{1,2}$, Wenyuan Gao$^{1}$, Xiaojie Zhou$^{1}$, \Letter Weijia Jia $^{2,1}$, Shaohua Zhang$^{1,2}$, Yutao Luo$^{1}$}
\address{$^{1}$Department of Computer Science and Engineering, Shanghai Jiao Tong University, Shanghai, China
\\$^{2}$State Key Lab of IoT for Smart City, CIS, University of Macau, Macau, SAR China
\\
sdq11111@sjtu.edu.cn, jiawj@um.edu.mo}

\maketitle

\begin{abstract}
Time-sync comment (TSC) is a new form of user-interaction review associated with real-time video contents,  which contains a user's preferences for videos and therefore well suited as the data source for video recommendations.  However, existing review-based recommendation methods ignore the context-dependent (generated by user-interaction), real-time, and time-sensitive properties of TSC data. To bridge the above gaps, in this paper, we use video images and users' TSCs to design an Image-Text Fusion model with a novel Herding Effect Attention mechanism (called ITF-HEA), which can predict users' favorite videos with model-based collaborative filtering. Specifically, in the HEA mechanism, we weight the context information based on the semantic similarities and time intervals between each TSC and its context, thereby considering influences of the herding effect in the model. Experiments show that ITF-HEA is on average 3.78\% higher than the state-of-the-art method upon F1-score in baselines.
\end{abstract}
\section{Introduction}
\label{intro}
Recently, watching online videos of news and amusement has become mainstream entertainment during people's leisure time. Therefore, efficient and accurate personalized video recommendation methods bring significant convenience to their life. Most of the video recommendation methods focus on users' behaviors such as their browsing history \cite{jannach2017recurrent,perera2017exploring} and reviews \cite{chen2017personalized,bauman2017aspect}. In real scenarios, however, most people are unwilling to do high-quality reviews after watching videos, which causes the scarcity of valuable video reviews. Furthermore, some multi-feature based methods \cite{gao2017unified,mei2011contextual} combine image information with review information to generate users' interests from a more comprehensive perspective. However, their methods have only achieved limited improvement because the images and reviews usually contain unequal information \cite{kiros2014multimodal}. That is, the text information and image information generally describe the different amount of contents. An image in a video only describes one moment of the video content, while a review usually describes the overall contents of the video. The information gap causes the fusion of reviews and images to lose great information. 

\begin{figure}[htbp]
  \centering
  \includegraphics[width=0.8\linewidth]{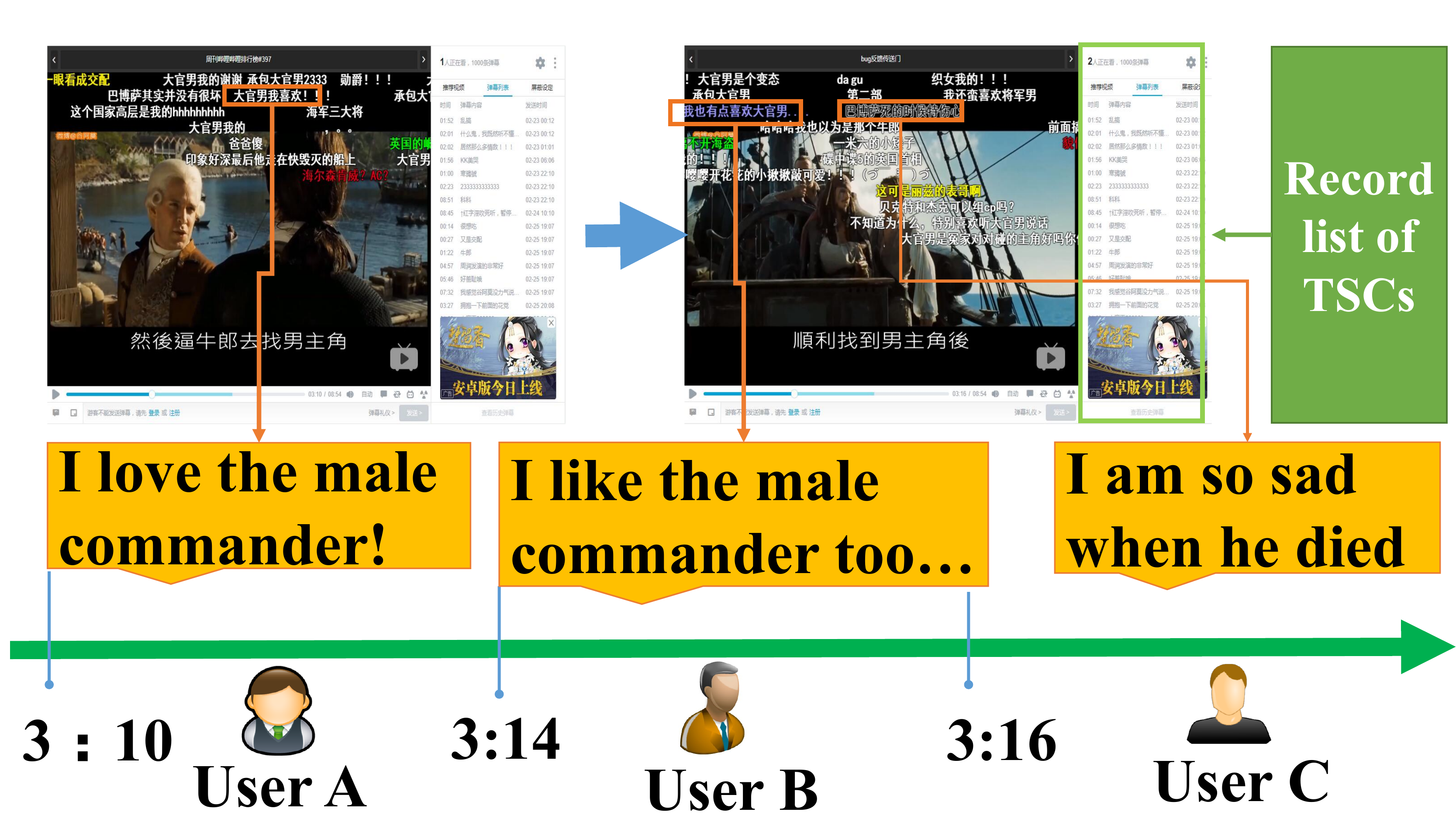}
  \caption{Examples of TSCs. }\label{fig:danmu}
\end{figure}

Meanwhile, a new form of user-interactive review -- time-sync comment (TSC) (first introduced by Wu \emph{et al.} \cite{wu2014crowdsourced}, see Fig. \ref{fig:danmu}) has become increasingly popular in China and Japan, especially among young people. Nowadays, many popular Chinese video websites such as Youku (http://www.youku.com), Bilibili (http://bilibili.tv) and the Japanese video website NICONICO (http://www.nicovideo.jp) support the TSC. TSCs convey information involving the content of the current video frame, feelings of users or replies to other TSCs, which can accurately express the users' preferences for the video. Moreover, each TSC has a corresponding timestamp to record the posted time. Compared with traditional video reviews, TSCs are much easier to obtain their corresponding images by timestamps. The users' real-time feedbacks and the vast amount make TSCs valuable and accessible sources for personalized video recommendations.

In this paper, we focus on mining the users' preferences and videos' features from TSCs and corresponding images to recommend videos towards users through model-based Collaborative Filtering (CF). TSCs have several features distinguished from the traditional video reviews: \textbf{\emph{(1)Context-dependent.}} TSCs are usually context-dependent, i.e., the latter comments often depend on the former ones. This phenomenon is known as the herding effect in social science \cite{he2016predicting,tchernichovski2017tradeoff}. An example of the herding effect is shown in Fig. \ref{fig:danmu}. User A said ''I love the male commander!'' to express his love to the role male commander when he appears in the video. After a few seconds, user B and user C followed up by saying "I like the male commander too..." and "I am so sad when he died."  In this case, user B and user C may not make their comments if user A has not. That is, the emergence of a TSC is usually not independent, but a probability event influenced by other preorder comments. \textbf{\emph{(2)Real-time.}}  Each TSC has a timestamp synchronous to the playback time of the video. The coverage of a TSC is usually only a short time before its timestamp. Therefore, the content of each TSC is closely related to the video content corresponding to its timestamp, which makes it easy to sample corresponding image information by timestamp. \textbf{\emph{(3)Time-sensitive.}} According to our observation, TSCs with a large interval of timestamps are unlikely to discuss the same topic, even if their semantics are similar. Users are more likely to follow those newer TSCs than older ones. As a result, the herding effect mentioned before will not last long. These features make TSC a particular review. However, most of the current TSC-based recommendations \cite{chen2017personalized,ping2018video} assume that TSCs are independent of each other and ignore the time information. Such assumption ignores the above features of TSCs which causes the loss of crucial semantic information and affects the accuracy of results. Therefore, how to take TSCs' features of context-dependent (herding effect), real-time and time-sensitive into account to extract the textual information and fuse it with visual information accurately and effectively are the central challenges.

Based on the above motivations and challenges, we propose an Image-Text Fusion model with a novel Herding Effect Attention mechanism (called ITF-HEA). In ITF-HEA, We generate users' preferences and summarize video contents through model-based CF. To analyze the influence of text information, image information and contextual information separately, we split ITF-HEA into two models:  Text-based Model(TM) and Image-Text Fusion model (ITF), and one attention mechanism: Herding Effect Attention (HEA) mechanism. Specifically, in TM, we sample and embed the TSCs to obtain the sentence vectors (TSC features) by bidirectional Long Short-Term Memory (LSTM) at first, and then combine TSC features with the hidden (embedding) features of the users and videos respectively to predict the likeness of the user to the video. In ITF, we sample the corresponding video frame (image) features and incorporate them with TSC features to replace single TSC features in TM. Finally, we design the HEA mechanism which is based on contextual semantic similarity and time interval of TSCs to incorporate contextual information into TSC features to replace the features in TM and ITF.

The main contributions of this paper are as follows:
\begin{itemize}
\item[1] We propose a novel HEA mechanism, which takes TSCs' features of contextual relevance, real time and timeliness into account, to extract the textual features of the TSCs more accurately and effectively.
\item[2] We design an Image-Text Fusion model using model-based CF, combine it with HEA mechanism and get the ITF-HEA,  which can predict the likeness of the user to the video more accurately and sufficiently.
\item[3] We evaluate the ITF-HEA with real-world datasets on mainstream video-sharing websites and compare it with state-of-the-art video recommendation methods. The results show that our model outperforms baselines in both precision and F1-score on average 3.3\% and 3.78\% respectively.
\end{itemize}

\section{Related work}
In this section, we discuss the related work in three aspects.

 \textbf{Time-sync video comments} are first introduced by Wu \emph{et al.} \cite{wu2014crowdsourced}. Then, Yang \emph{et al.} \cite{yang2017crowdsourced} sum up the features of TSCs, which inspire our work. LV \emph{et al.} \cite{lv2016reading} propose a video understanding framework to assign temporal labels to highlighted video shots. They are the first to analyze the TSCs using the neural network. Recently, Liao \emph{et al.} \cite{liao2018tscset} present a larger-scale TSC dataset with four-level structures and rich self-labeled attributes, which brings convenience for future research on TSCs. The above methods show that TSC is a kind of data with great potential and development value.

 \textbf{Video recommendation} has attracted great attention from both the industry and academia. Most of the current state-of-the-art methods are based on CF. Mcauley \emph{et al.} \cite{mcauley2013hidden} combine latent rating dimensions with latent review topics, which is a review-based method. Diao \emph{et al.} \cite{diao2014jointly} propose a probabilistic model based on CF and topic modeling, which is an LDA \cite{blei2003latent} based method and allows capturing the interest distribution of users and the content distribution of movies. He \emph{et al.} \cite{he2016vbpr} propose a scalable factorization model to incorporate visual signals into predictors of people's opinions, which is the state-of-the-art visual-based model. However, the above recommendation methods are not well-designed for TSCs as they ignore the interactive, real-time, and timeliness properties of TSC data.

 \textbf{Attention Mechanism} has been shown effective in natural language processing \cite{duan2018attention,qian2018translating,liu2018neural}. Recently, attention models have been used increasingly in recommendation systems to assign weights to user-item pairs. Chen \emph{et al.} \cite{chen2017attentive} introduce a novel attention mechanism in CF to address the challenging item and component-level implicit feedback in the multimedia recommendation, which can be seamlessly incorporated into classic CF models with implicit feedback.  Seo \emph{et al.} \cite{seo2017interpretable} propose to model user preferences and item properties using convolutional neural networks (CNN) with dual local and global attention. Our herding effect attention mechanism adopts the soft attention \cite{vaswani2017attention}, which learns the attentive weights based on the importance to the final task.

\section{Model-based Collaborative Filtering}
In this section, we describe two  CF models and an attention mechanism. First, the problem formulation is provided in Section \ref{4.0}. Then, we propose a Text-based Model by using textual features of TSCs in Section \ref{4.1}. Next, we design an Image-Text Fusion Model to jointly model video images as well as TSCs in Section \ref{4.2}. Finally, to take full consideration of the features of TSCs, we implement Herding Effect Attention mechanism and give the complete neural network structure of the Image-Text Fusion Model with Herding Effect Attention in Section \ref{4.3}.
\label{4}
\subsection{Problem Formulation}
\label{4.0}

Suppose there are $N$ TSCs, $\boldsymbol{TSC}$ $=$ $\{tsc_{1}, tsc_{2}, ...,tsc_N\}$. For $tsc_{i}$, we define the corresponding visual feature as $\boldsymbol{vsl_{i}}$ (see Section \ref{4.2} for details) and sentiment polarity $pol_{i}$ which is determined by the Stanford sentiment analysis toolkit (http://nlp.stanford.edu/sentiment). Besides, we define $u_i$ to represent the user ID and $v_i$ to express the video ID of $tsc_{i}$.

As mentioned in section \ref{intro}, TSCs are easily affected by previous comments. Therefore, for $tsc_{i}$, we continuously sample M preorder TSCs $\boldsymbol{Context_{i}}=\{pre_{i,1},pre_{i,2},...,pre_{i,M}\}$ as context information ($pre_{i,M}$ is $tsc_i$ itself). For each $pre_{i,j} \in \boldsymbol{Context_i}$, we define the time-stamp $t_{i,j}$ to represent its posted video time. The word list of $tsc_i$ and its context information $pre_{i,j}$ are defined as $\boldsymbol{w_{i}}=\{w_{i}^{1},w_{i}^{2},...,w_{i}^{L_{i}}\}$ and $\boldsymbol{w_{i,j}}=\{w_{i,j}^{1},w_{i,j}^{2},...,w_{i,j}^{L_{i,j}}\}$ where $L_{i}$ and $L_{i,j}$ are the length of $tsc_i$ and $pre_{i,j}$.

Given total TSCs $\boldsymbol{WT}$ $=$ $\{\boldsymbol{w_{i}}| 1 \leq i \leq N \}$ and their corresponding sentiments $\boldsymbol{POL}$ $=$ $\{pol_{i}|1\leq i\leq n \}$, user ID list $\boldsymbol{U}$ $=$ $\{u_1,u_2,...,u_N\}$, video ID list $\boldsymbol{V}$ $=$ $\{v_1,v_2,...,v_N\}$, visual feature list $\boldsymbol{VSL}$ $=$ $\{vsl_1,vsl_2,...,vsl_N\}$, and the context information 
$
\boldsymbol{WC}$ $=$ 
$
\{
\{\boldsymbol{w_{1,1},...,w_{1,M}}\},
...,
\{\boldsymbol{w_{N,1},...,w_{N,M}}\}
\}
$ with the corresponding time-stamp
$
\boldsymbol{T}$ $=$
$
\{
\{t_{1,1},...,t_{1,M}\},
...,
\{t_{N,1},...,t_{N,M}\}
\}
$, our task is to predict the likeness of user $u\in \boldsymbol{UID}$ on video $v\in \boldsymbol{VID}$, where $\boldsymbol{UID}$ and $\boldsymbol{VID}$ are the sets that contain the unique IDs in $\boldsymbol{U}$ and $\boldsymbol{V}$.
The top $X$ videos among the final results are recommended to the corresponding user. 

\subsection{Text-based Model}
\label{4.1}
\begin{figure}[htbp]
  \centering
  \includegraphics[width=0.85\linewidth]{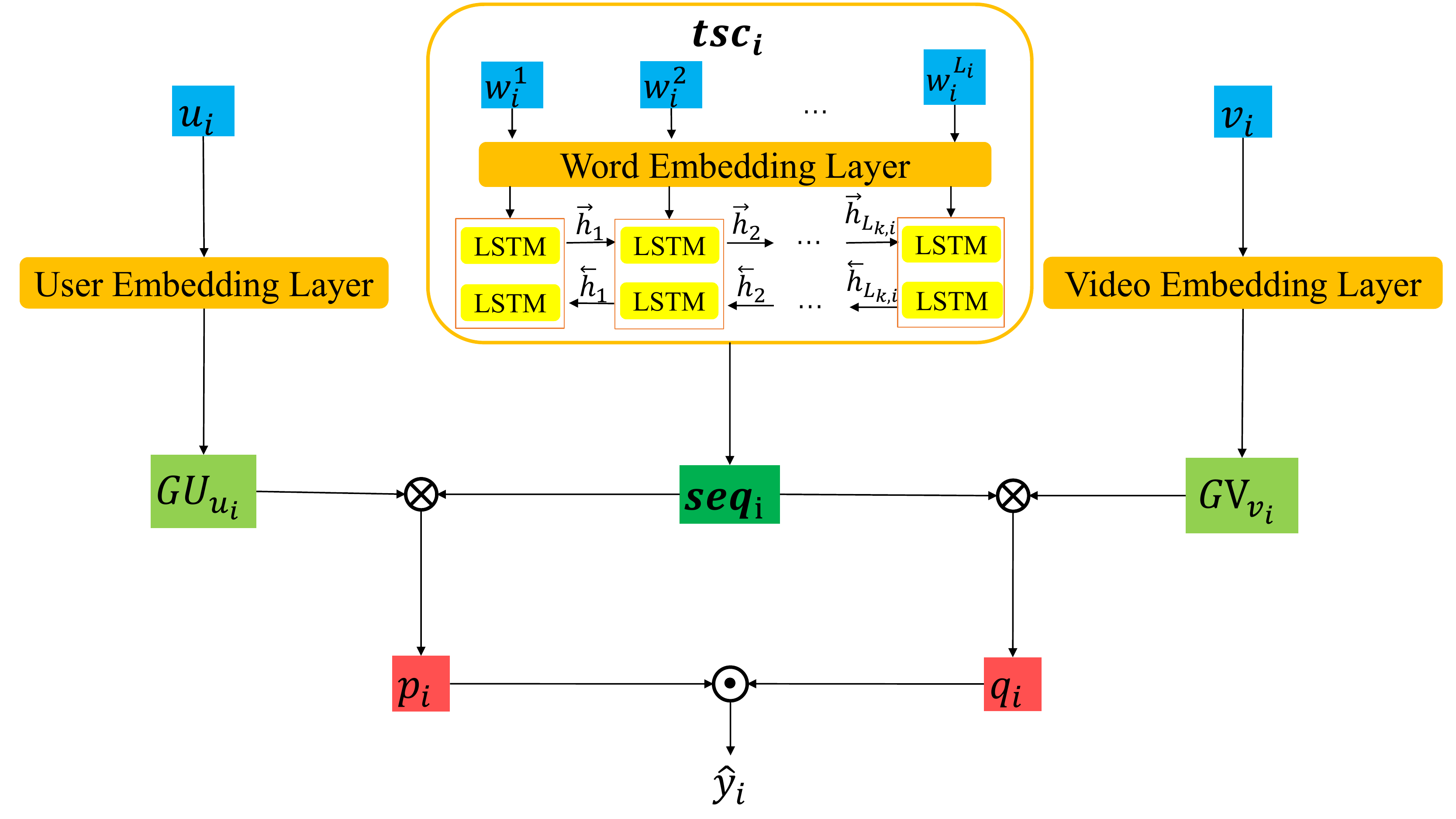}
  \caption{Text-based Model}\label{fig:gen}
\end{figure}
Intuitively, the preference of users is extracted from their published TSCs in corresponding videos, while the textual features of the videos are summed up from the TSCs published in the video. Based on above, in this section, we first extract the features of TSCs by bidirectional LSTM. Then, features of TSCs are merged with the latent factors of users and videos. Finally the likenesses of the users to videos are predicted by CF. The general framework of Text-based Model (TM) is shown in Fig. \ref{fig:gen}.

More concretely, to capture the word sequential information from TSCs, we use the Bidirectional Long Short-Term Memory (Bi-LSTM) network \cite{schuster1997bidirectional} to convert word features into TSC features.

For each $tsc_i$, we have 
\begin{eqnarray}
&&\boldsymbol{ \overrightarrow{h} _{t}}= LSTM( {w_{i}^{t}}, \boldsymbol{\overrightarrow{h}_{t-1}}) \label{first}\\
&& \boldsymbol{ \overleftarrow{h} _{t}}= LSTM( {w_{i}^{t}}, \boldsymbol{\overleftarrow{h}_{t+1}})
\end{eqnarray}
and
\begin{equation}
\label{seq}
\boldsymbol{seq_i}=\frac{1}{L_i}\sum_{t=1}^{L_i} (\overrightarrow{\boldsymbol{h} _{t}} \oplus \overleftarrow{\boldsymbol{h}_{t}})
\end{equation}
where $\boldsymbol{seq_i} \in \mathbb{R}^d$ and $\oplus$ denotes vector concatenation.
After LSTM layer, we get the sequence feature $\boldsymbol{seq_{i}}$ as the output. 

Then, we define $\boldsymbol{GU_{u_i}}$ as the latent factor of user $u_i$, which is the feature based on user's historical preference. Likewise, the feature of video $v_i$ is defined as
$\boldsymbol{GV_{v_{i}}}$. Afterward, we design $\otimes$ function to merge $\boldsymbol{GU_{u_{i}}}$ and $\boldsymbol{GV_{v_{i}}}$ with $\boldsymbol{seq_{i}}$ respectively, and obtain 
\begin{equation}
\boldsymbol{p_{i}}=\boldsymbol{G_{u_{i}}} \otimes \boldsymbol{seq_{i}}
\label{eq:merge1}
\end{equation}
and
\begin{equation}
\boldsymbol{q_{i}}=\boldsymbol{G_{v_{i}}} \otimes \boldsymbol{seq_{i}}
\label{eq:merge2}
\end{equation}
 where $\otimes$:$R^{d} \times R^{d} -> R^{d}$ is an element-wise product function to merge two $d$ dimensional vectors into one. Specifically, $$(a_{1}, ..., a_{d}) \otimes (b_{1}, ..., b_{d})=(a_{1}b_{1}, ..., a_{d}b_{d})$$

In our framework, we take the prediction of a user's favor to a video (or video clip) as a binary classification problem, where 1 means a user likes the video, and 0 otherwise. Therefore, we define the likeness of user $u_i$ to video $v_i$ though $tsc_i$ in the training data as 
\begin{equation}
\hat{y}_{i}= sigmoid(\boldsymbol{p_{i}  \odot q_{i}})
\label{eq:pred}
\end{equation}
where $sigmoid(x)=\frac{1}{1+e^{-x}}$ and ``$\odot$'' denotes inner product.  

Generally, users comment on their favorite videos with positive sentiment. Therefore, we determine the polarity of each TSC by the Stanford sentiment analysis toolkit \cite{manning}. For simplicity, we set the polarity of each TSC as 1 if the result is positive or neutral, and 0 otherwise. We define 
$
y_{i}= pol_{i}
$
as the ground truth of the likeness of user $u_i$ for video $v_i$ through $tsc_i$, where $pol_{i}$ is the polarity of $tsc_i$.

At last, we use the binary cross-entropy as our loss function to model user preference. The final objective function is maximized as:
\begin{equation}
\begin{split}
L = & \sum_{i=1}^{N}(y_{i}\cdot ln \hat{y}_{i} + (1-y_{i}) \cdot ln(1-\hat{y_{i}}))
\end{split}
\end{equation}
In the training phase, the parameters can be learned via Adam \cite{kingma2014adam}.

After trainning, we use 
\begin{equation}
\label{eq:prep}
\hat{y}_{u,v}=\boldsymbol{ GU_{u}}\odot \boldsymbol{ GV_{v}} 
\end{equation} 
to express the predicted likeness of user $u$ on video $v$, and 
\begin{equation}
Po_{u,v}=\frac{\sum_{i \in List_{u,v}}{pol_{i}}}{|List_{u,v}|}
\end{equation}
to express the real likeness of user $u$ for video $v$, where $List_{u,v}$ expresses the total TSCs that user $u$ has commented on video $v$.
Then, the ground truth of testing data is defined as 
\begin{equation}
y_{u,v}=
\begin{cases}
0 & Po_{u,v} < 0.5 \\
1 & Po_{u,v} \geq 0.5
\end{cases}
\end{equation} 

\subsection{Image-Text Fusion Model}
\label{4.2}
\begin{figure}[htbp]
  \centering
  \includegraphics[width=0.85\linewidth]{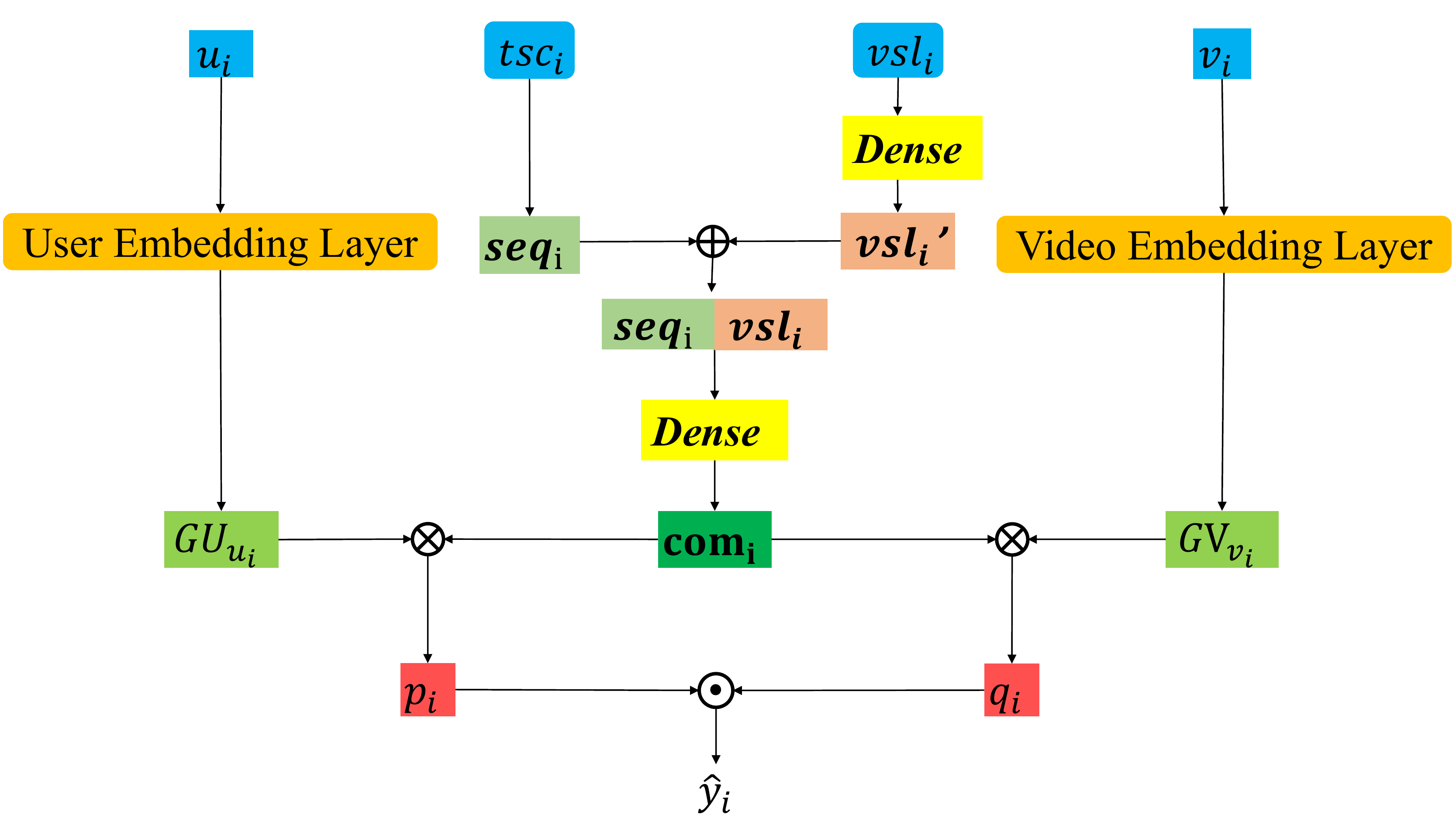}
  \caption{Image-Text Fusion Model}\label{fig:com}
\end{figure}
In the time-sync video, each TSC has a timestamp that records the corresponding video time when the TSC is published. So that, we can easily obtain the corresponding image information for better feature extraction. In this section, we focus on merging TSC text features with corresponding visual features to obtain more comprehensive features.
The general framework of Image-Text Fusion model (ITF) is shown in Fig. \ref{fig:com}.

For $tsc_i$, we use $vsl_{i}$ to indicate the visual feature when $tsc_i$ is posted. The visual features are with the output of 4096-way obtained from a public TSC data set extracted by \emph{Chen et al.} \cite{chen2017personalized}, which are trained by the Caffe reference model with 5 convolutional layers followed by 3
fully-connected layers that have been pre-trained on 1.2 million ImageNet (ILSVRC2010) images.

Since the dimension of $\boldsymbol{vsl_{i}}$ is 4096, we reduce its dimension to $d$ and get $\boldsymbol{vsl_{i}'}$ by 
\begin{equation}
\boldsymbol{vsl_i'}=Dense(\boldsymbol{vsl_i})
\end{equation} 
where $Dense$ is the fully-connected layer with the activation function $elu$ \cite{clevert2015fast}.

To combine the image features and textual features, we first concatenate the sequence feature $seq_{i}$ with the visual feature $vsl_{i}'$, and obtain the $2\times{d}$-dimensional vector $com_{i}$:
\begin{equation}
\boldsymbol{com_{i}} = \boldsymbol{seq_{i}} \oplus \boldsymbol{vsl_{i}'}
\end{equation} 
Then, we reduce the demension of $com_{i}$ to $d$ and get the $com_{i}'$ by 
\begin{equation}
\boldsymbol{com_{i}'} = Dense(\boldsymbol{com_{i}})
\end{equation} 

Finally, we use $com_{i}'$ instead of $seq_{i}$ to merge with $\boldsymbol{GU_{u_{i}}}$ and $\boldsymbol{GV_{v_{i}}}$ by $Eq. (\ref{eq:merge1})$ and $Eq. (\ref{eq:merge2})$ and predict the likeness by $Eq. (\ref{eq:pred})$.

\subsection{Herding Effect Attention}
\label{4.3}
\begin{figure}[htbp]
  \centering
  \includegraphics[width=0.85\linewidth]{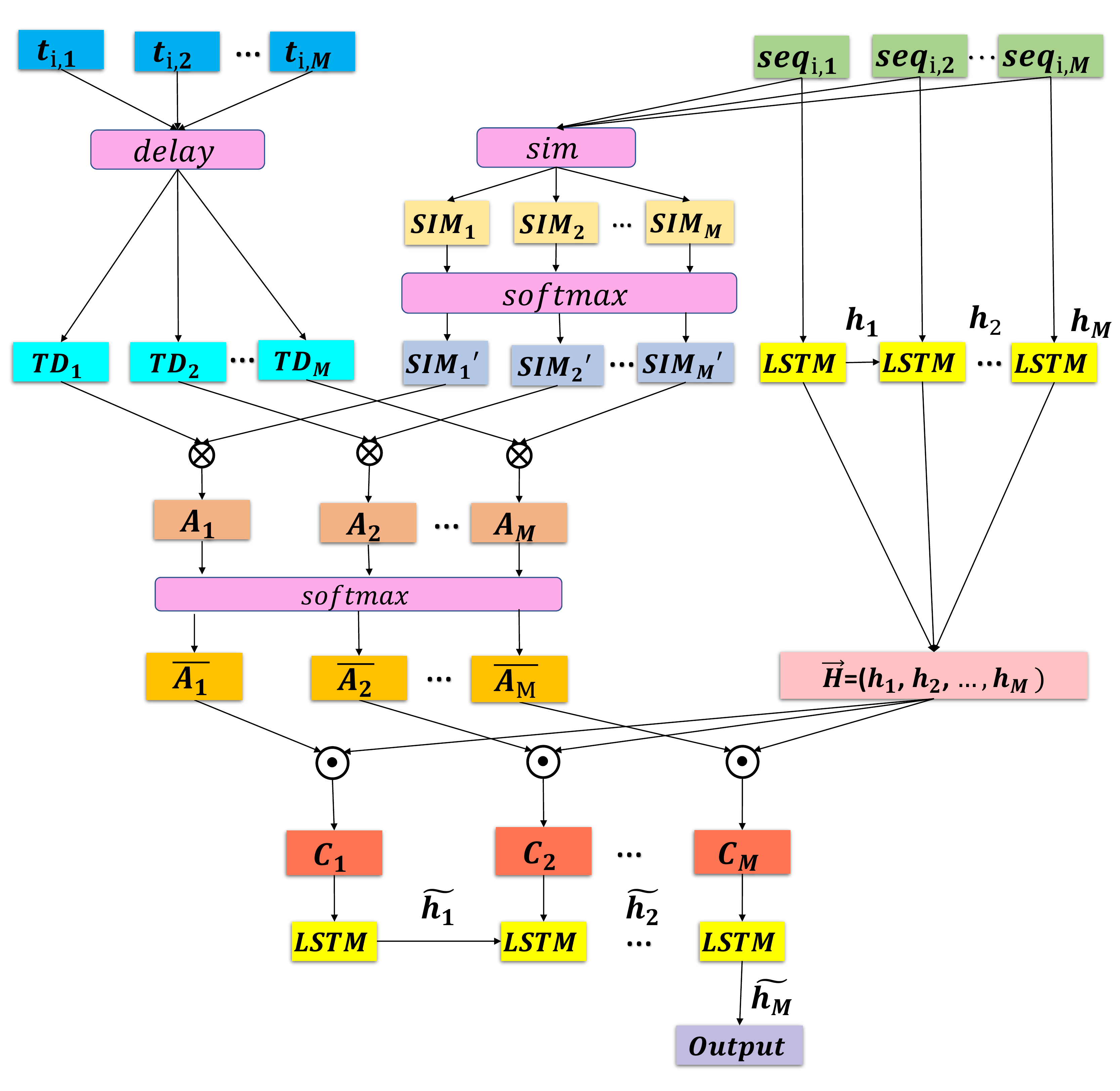}
  \caption{Herding Effect Attention}\label{fig:attention}
\end{figure}
Existing review-based recommendation methods usually handle each comment separately without considering the context associations between the comments. However, TSCs are highly semantic relevant and time-related, which is so-called the herding effect. That is, TSCs may be affected by other preorder TSCs on the similar topic. Also, TSCs with similar semantics and the short interval of the time-stamp are more likely to influence each other. Based on above, we design an HEA mechanism, which calculates the influence weights of TSC contexts by their semantic similarities and timestamp intervals in an LSTM-based encoder-decoder framework. The framework of HEA is shown in Fig. \ref{fig:attention}.

Formally, for $tsc_i$, we sample $M$ continuous preorder TSCs  $\boldsymbol{Context_{i}}=\{pre_{i,1},pre_{i,2},...,pre_{i,M}\}$ as the context information and get context features $\boldsymbol{SEQ_i}=[\boldsymbol{seq_{i,1}, ..., seq_{i,M}} ]$ by $Eq. (\ref{first})$ - $(\ref{seq})$, where $t_{i,1}<t_{i,2}<....<t_{i,M}$.

We formalize the HEA into an encoder-decoder framework. Given context features $\boldsymbol{SEQ_i}$ as the input to the LSTM, the output are obtained as:
\begin{eqnarray}
&& \boldsymbol{h_{t}}=LSTM( \boldsymbol{h_{t-1}}, \boldsymbol{seq_{i,t}})
\end{eqnarray}
We use $\boldsymbol{H}=(\boldsymbol{h_1,h_2,...,h_M})$ to express the hidden status vectors of the encoder output.

To calculate the influence weights of contexts of TSCs, for $pre_{i,j}$, we define the semantic similarity vector and time delay vector as $\boldsymbol{SIM_{j}}=(sim(j,1),sim(j,2),...,sim(j,M))$ and $\boldsymbol{TD_{j}}=(delay(j,1),delay(j,2),...,delay(j,M))$ , where 
\begin{equation}
sim(k,j)= \frac{\boldsymbol{seq_{i,k}}\odot \boldsymbol{seq_{i,j}}}{|\boldsymbol{seq_{i,k}}||\boldsymbol{seq_{i,j}}|}
\end{equation}
represents the semantic similarity between $pre_{i,j}$ and $pre_{i,k}$, and 
\begin{equation}
delay(j,k)=
\begin{cases}
\mathrm{e}^{-\beta(t_{i,j}-t_{i,k})}    & j>k\\
0 & j<=k
\end{cases}
\end{equation} represents the influence of $pre_{i,k}$ on $pre_{i,j}$ decreases with the increasing time interval ($\beta$ is a hyper-parameter that will be discussed in Section \ref{5}).

Since the semantic similarity may have negative numbers, we first normalize $\boldsymbol{SIM_{j}}$ by softmax as: $$\boldsymbol{SIM_{j}}'=(\frac{\mathrm{e}^{sim(j,1)}}{\sum_{k=1}^{M}\mathrm{e}^{sim_{j,k}}}, \frac{\mathrm{e}^{sim(j,2)}}{\sum_{k=1}^{M}\mathrm{e}^{sim_{j,k}}}, ..., \frac{\mathrm{e}^{sim(j,M)}}{\sum_{k=1}^{M}\mathrm{e}^{sim_{j,k}}})$$

Next, we calculate the attention score vector of $pre_{i,j}$ as:
\begin{equation}
\boldsymbol{A_{j}}= \boldsymbol{SIM_{j}}' \otimes \boldsymbol{TD_{j}}
\end{equation}

The final attention score distribution $\boldsymbol{\overline{A}_j}$ is obtained by normalizing the attention score vector $\boldsymbol{A_{j}}$ by softmax function.

We compute the input of decoder as:
\begin{equation}
\boldsymbol{C_{j}}=  \boldsymbol{\overline{A}_j}  \odot \boldsymbol{H}
\end{equation}
and get the output as:
\begin{equation}
\boldsymbol{\widetilde{h}_{1}}=LSTM(\boldsymbol{C_{1}})
\end{equation}
\begin{equation}
\boldsymbol{\widetilde{h}_{t}}=LSTM(\boldsymbol{C_{t}}, \boldsymbol{\widetilde{h}_{t-1}})
\end{equation}
Finally, we use $\boldsymbol{\widetilde h_M}$ instead of $\boldsymbol{seq_{i}}$ that we used in Section \ref{4.1} and \ref{4.2} as the textual feature. 

\begin{figure}[htbp]
  \centering
  \includegraphics[width=0.85\linewidth]{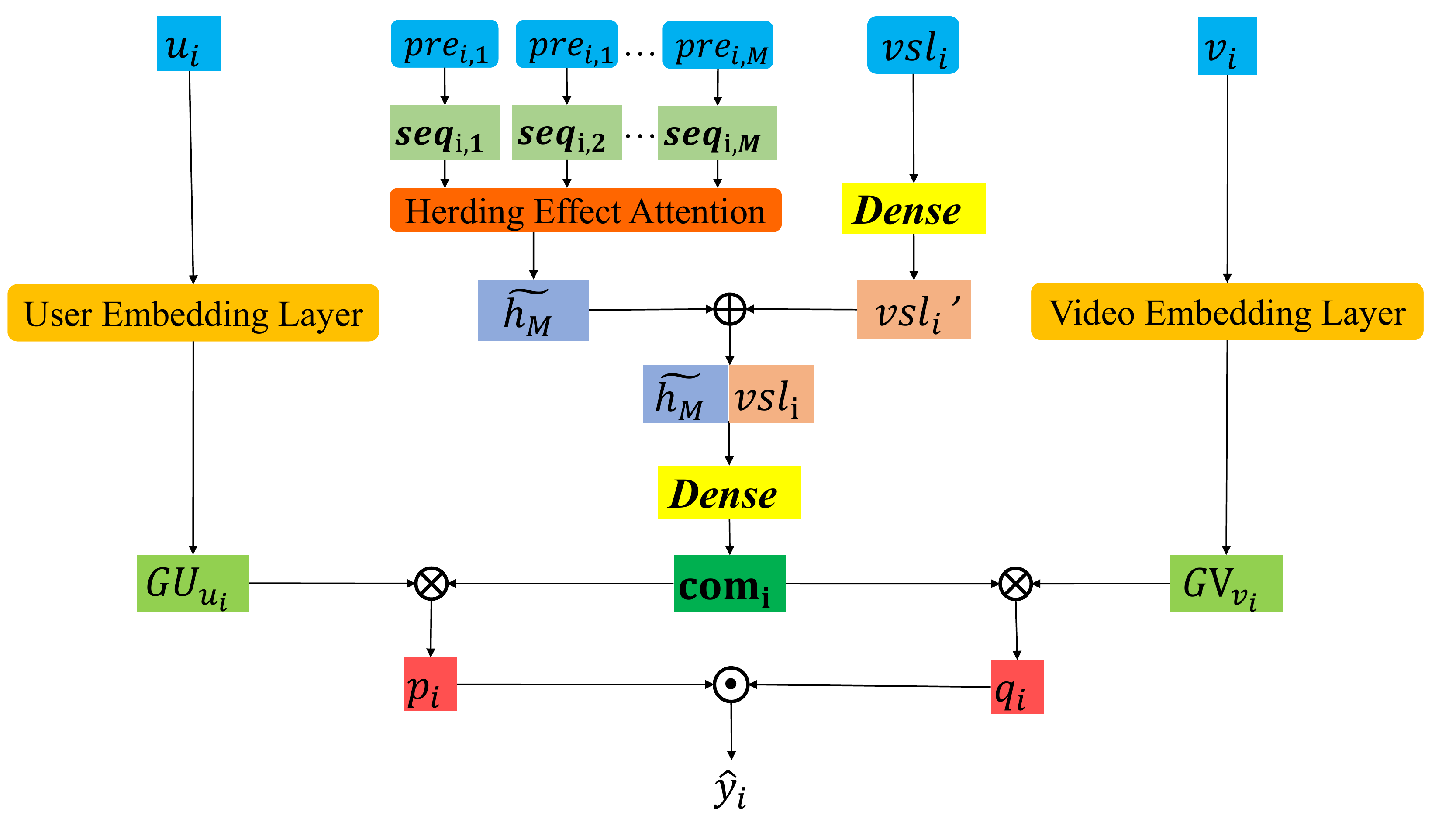}
  \caption{Complete network structure}\label{fig:att_com}
\end{figure}

We integrate the context-dependent, real-time and time-sensitive properties of the TSCs into the model by the HEA mechanism, which can be applied in both TM and ITF to improve the TSC feature extraction.
The complete network structure of the ITF-HEA is shown in Fig. \ref{fig:att_com}.

\section{Experiments}
\label{5}
In this section, we demonstrate the effectiveness of our proposed method by comparing with 4 well-known methods of video recommendation. We provide necessary parameters of our model at first and then analyze the performance of our model on time-sync video recommendation. Finally, we analyze the effect of the hyper-parameters on the experimental results.

\subsection{Experimental Setup and Dataset}
\label{set}

The data used in this paper are crawled from a Chinese time-sync video site Bilibili by Chen \emph{et al.} \cite{chen2017personalized}, which are obtained from the movie category till December 10th, 2015. In this paper, we select 100 users who have posted the most TSCs and commented on more than 40 videos. These users have commented on a total of 871 videos, and we select all the comments in those videos as a sub-dataset. In the sub-dataset, 423,384 users have published 1,319,475 TSCs in total.

For each of the 100 users, we select half of the videos where they have commented as the training set, and the other half as the test set. We make sure that at least 20 videos per user can be recommended (To ensure the effectiveness of top20). In the test set, we get 2,995 $(user, video)$ pairs, where 1,972 pairs are positive, and 1,023 pairs are negative in sentiment polarity. 

In the training set, we obtain 2,811 $(user, video)$ pairs with 11,775 TSCs (a user may make more than one TSC in a video), where 8,124 TSCs are positive and 3,651 TSCs are negative.

In our model, hyper-parameter $\beta$ and number of contextual TSCs $M$ need to be decided. We select 35\% data of the test set (actually 1,075 $(user,video)$ pairs) as the validation set to tune $\beta$. The initial learning rate of Adam \cite{kingma2014adam} is 0.001 and  the vector dimension $d$ is set as 128. We get the best results when  $\beta=0.2$, and $M=10$, which are discussed in Section  \ref{result}.

\subsection{Results}
\label{result}
In this section, we use the test set described in Section \ref{set} to compare our complete model with existing methods. 

To evaluate the performance of the proposed models, we compare our model with the following methods as baselines:
\begin{itemize}

\item \textbf{HFT:} A state-of-the-art method regarding making rating prediction with textual reviews \cite{mcauley2013hidden}. In the experiments, we set the ratings of positively commented videos as 1, and 0 otherwise.

\item \textbf{JMARS:} A Latent Dirichlet Allocation (LDA) based method to make rating prediction with textual reviews \cite{diao2014jointly}.

\item \textbf{VBPR:} A visual-based recommendation method \cite{he2016vbpr}.

\item \textbf{KFRCI:} A novel Key Frame Recommender by modeling user TSCs and keyframe Images simultaneously \cite{chen2017personalized}. In the experiments, the likeness score of the video is the average score of all the frames the users have commented on.

\item \textbf{ITF-HEA:} The Image-Text Fusion Model proposed in Section \ref{4.2} with Herding Effect Attention mechanism proposed in Section \ref{4.3}.

\end{itemize}

\begin{table}[!htbp] 
\caption{Precision and F1-score of each method} \label{table:pre}
\footnotesize
\center
\begin{tabular}{|c|c|c|c|c|c|c|}
\hline
&\multicolumn{2}{|c|}{Top5 }& \multicolumn{2}{|c|}{Top10 }& \multicolumn{2}{|c|}{Top20} \\

& Prec & F1 & Prec & F1 & Prec & F1\\
\hline
HFK & 0.856& 0.3463& 0.812 &0.5310& 0.732& 0.7371\\
\hline
JMARS & 0.878 &0.3552& 0.810 &0.5560& 0.732 &0.7367\\
\hline
VBPR & 0.892 &0.3608& 0.83 &0.5760& 0.779 &0.7840\\
\hline
KFRCI & 0.954 &0.3859& 0.897 &0.6036& 0.818&0.8233\\
\hline
\textbf{ITF-HEA} & \textbf{0.976}&\textbf{0.3948}& \textbf{0.932} &\textbf{0.6272}& \textbf{0.860}&\textbf{0.8656}\\
\hline
\end{tabular}
\end{table}

For each method in the baselines, we select a set of the best experimental parameters according to the range of the parameters given in their experiments and calculate the likenesses/ratings between users and videos.

Our experiments are conducted by predicting Top 5, 10, and 20 favorite videos respectively. The Top $X$ is the top prediction of user's likeness to the videos in test set calculated by Eq. (\ref{eq:prep}). We recommend all X videos to each user and consider these are the user's favorite videos.
We adopt F1-score and precision to evaluate the performance of the baselines and our models. All the models are repeated for 10 times, and we report the average values as the final results for clear illustration.

The results of F1-score and precision are shown in Table \ref{table:pre}. From Table \ref{table:pre}, we can see: ITF-HEA achieves the best performance on F1 and precision (F1 is proportional to precision in the Top 5, 10 and 20). It has enhanced the performance by about $2.30\%$, $3.91\%$ and $5.14\%$ (on average 3.78\%) upon F1-score and 2.20\%, 3.50\% and 4.20\% (on average 3.30\%) upon Precision on Top5, 10 and 20 respectively compared with KFRCI, which performs best among baselines. In other methods of baselines, the vision-based method VBPR has better performance than the others; the text-based method HFT and JMARS have similar performance, while the PMF method has the worst. 

Next, we compare the models proposed in Section \ref{4}: 
\begin{itemize}
\item \textbf{TM:} The Text-based Model proposed in Section \ref{4.1}.

\item \textbf{T-HEA:} The Text-based Model proposed in Section \ref{4.1} with Herding Effect Attention mechanism proposed in Section \ref{4.3}.

\item \textbf{ITF:} The Image-Text Fusion Model proposed in Section \ref{4.2}.

\item \textbf{ITF-HEA:} The Image-Text Fusion Model proposed in Section \ref{4.2} with Herding Effect Attention mechanism proposed in Section \ref{4.3}.
\end{itemize}
to analyze the effects of text features, image features and the attention mechanism in our model. The results of F1-score and precision are shown in Table \ref{table:self}.

\begin{table}[!htbp] 
\caption{Precision and F1-score of the models proposed in Section \ref{4}} 
\footnotesize
\label{table:self}
\center
\begin{tabular}{|c|c|c|c|c|c|c|}
\hline
&\multicolumn{2}{|c|}{Top5 }& \multicolumn{2}{|c|}{Top10 }& \multicolumn{2}{|c|}{Top20} \\
& Prec & F1 & Prec & F1 & Prec & F1\\
\hline
TM &0.932&  0.2363&  0.887 &0.5969& 0.790 &0.7950\\
\hline
T-HEA & 0.956 &0.3867& 0.914& 0.6151& 0.817 &0.8223\\
\hline
ITF & 0.952& 0.3851& 0.892 &0.6003& 0.805& 0.8107\\
\hline
\textbf{ITF-HEA} & \textbf{0.976} & \textbf{0.3948} & \textbf{0.932}  & \textbf{0.6272} & \textbf{0.860} & \textbf{0.8656}\\
\hline
\end{tabular}
\end{table}

The results show that although T-HEA only uses the textual information, the experimental results are still better than ITF. T-HEA even has better performance than the state-of-the-art method KFRCI, which shows that our HEA mechanism can effectively offset the effects of the herding effect and improve the performance of the model. The results also show that the context and timestamp of the TSCs are vital information and need to be considered.

\begin{figure}[!htbp]
\centering
\begin{minipage}[t]{0.48\linewidth}
\centering
\includegraphics[width=1\linewidth]{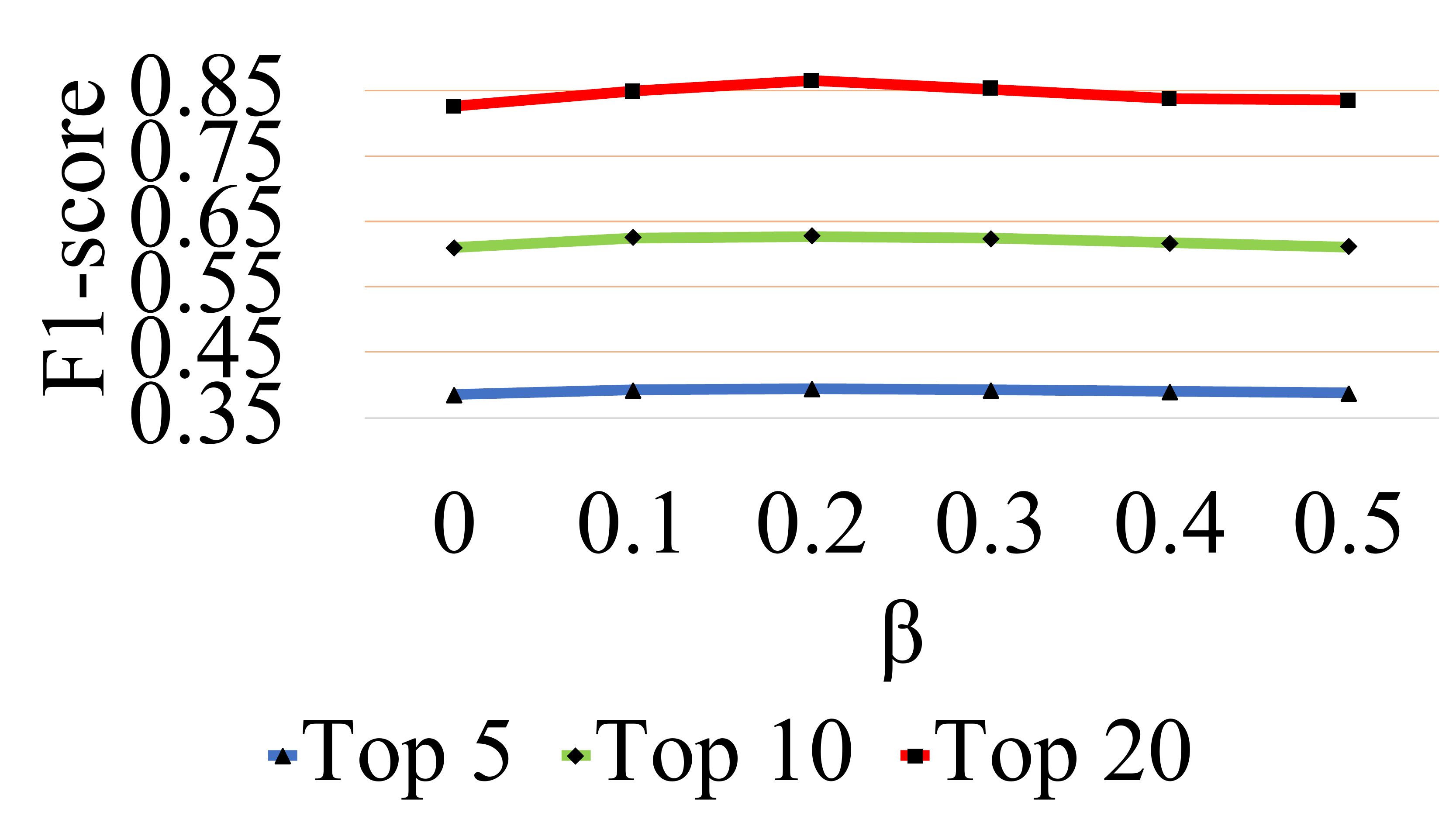}
\caption{The influence of the hyper-parameter $\beta$ on Top 5,  Top 10 and Top 20}
\label{fig:beta}
\end{minipage}
\begin{minipage}[t]{0.48\linewidth}
\centering
\includegraphics[width=1\linewidth]{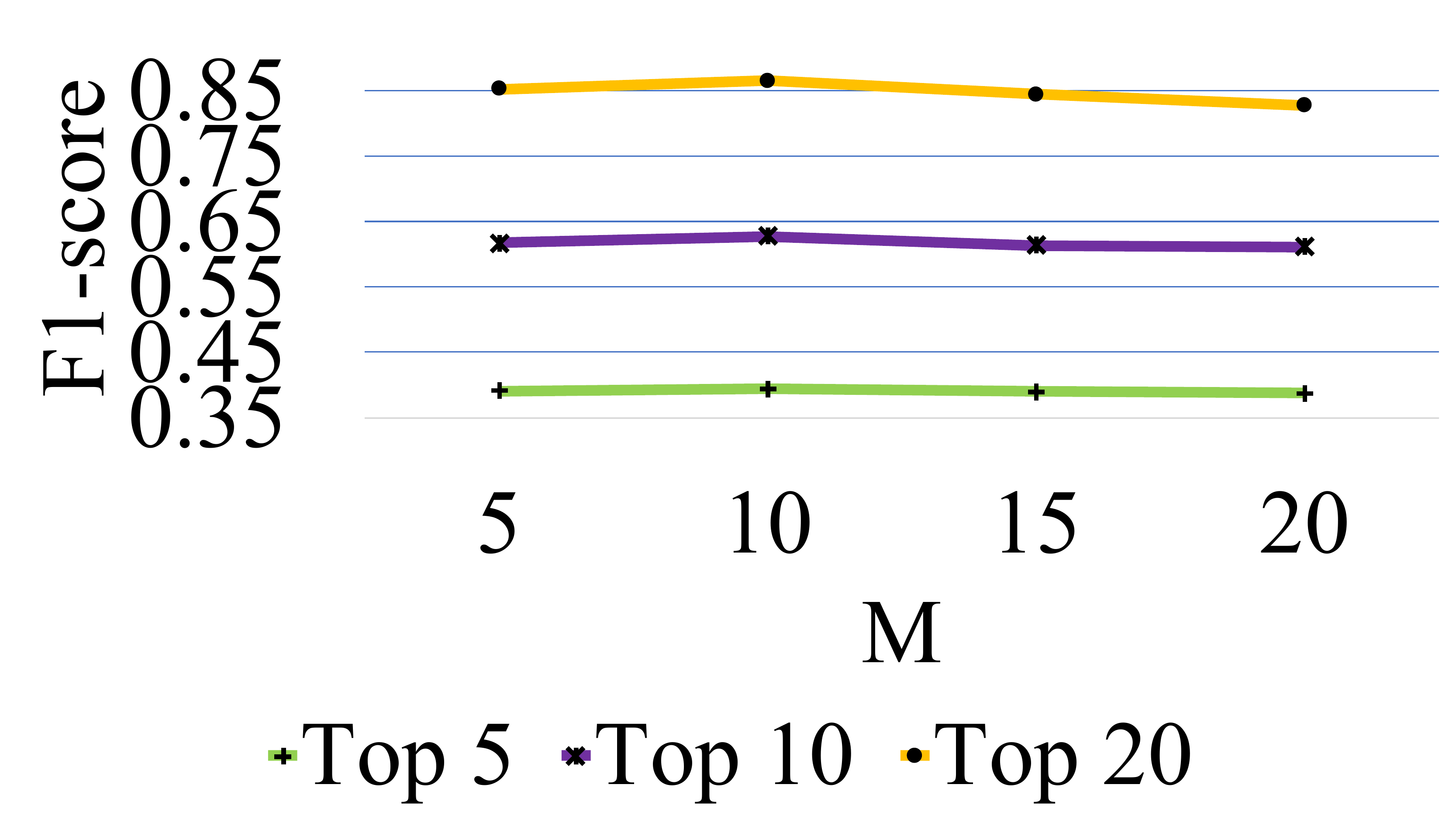}
\caption{The influence of the number of context $M$ on Top 5,  Top 10 and Top 20}
\label{fig:M}
\end{minipage}
\end{figure}
Finally, we discuss the influence of hyper-parameter $\beta$ and the number of contextual TSCs $M$ on the experimental results. We fix $M=10$, changing the value of $\beta$ from 0 to 0.5 (the step size is 0.1), and calculate the F1-score of Top 5,10 and 20 users' favorite videos in the validation set at first. The best results are obtained when $\beta=0.2$. We also calculate the F1-score for the different hyper-parameters in the test set, and the results are shown in Fig. \ref{fig:beta}. The hyper-parameter $\beta$ gains the best performance when $\beta=0.2$ in any case, which is the same with the validation set. When $\beta$ is bigger, the result of the experiment is worse because it weakens the weight of other TSCs in the attention layer.  When $\beta=0$, it has the worst performance, because the time information is not considered. 

For the number of context length $M$, we fix $\beta=0.2$, and set $M$ as 5, 10, 15 and 20, respectively. The results are shown in Fig. \ref{fig:M}. IFT-HEA gets the best performance when $M=10$ and the worst when $M=20$. This result confirms that the herding effect of TSCs is time-sensitive and will not last long, which meets our observation in Section \ref{intro}.

\section{ Conclusion and Future Work}
In this paper, we proposed a novel personalized online video recommendation with the dataset of both TSCs and its corresponding images through model-based CF method. To extract the textual features of the TSCs more accurately and effectively, we designed the HEA mechanism to add influence weight to each TSC based on their semantic similarity and time interval. In this way, we integrated the context-dependent, real-time and time-sensitive properties of TSCs in the neural network framework and predicted the users' preferences for online videos accurately and effectively. Extensive experiments on real-world dataset proved that our model could recommend videos to users more precisely than the state-of-the-art method with the HEA mechanism.

This is the first step towards our goal in personalized video recommendation, and there is much space for further improvements. For example, to design a more accurate fusion model to capture comprehensive user preference is a challenging problem. Besides, how to measure the weight of each video to the user's preference is also a challenging problem.

\section{Acknowledgements}
This work is supported by National China 973 Project No. 2015CB352401; Chinese National Research Fund (NSFC) Key Project No. 61532013 and No. 61872239. 0007/2018/A1, DCT-MoST Joint-project No. 025/2015/AMJ,FDCT,SAR Macau, China, and University of Macau Grant Nos: MYRG2018-00237-RTO, CPG2019-00004-FST and SRG2018-00111-FST.

\bibliographystyle{IEEEbib}
\bibliography{icme2019template}
\end{spacing}

\end{document}